\documentclass[aps,a4paper,12pt, twocolumns]{revtex4}
\usepackage{float}
\usepackage{fancyhdr}
\usepackage{color}
\usepackage{graphicx}
\usepackage{epsfig} 
\usepackage{amssymb}
\usepackage{amsmath,amsfonts}
\usepackage{booktabs}
\usepackage{hyperref}
\usepackage{pgfplotstable}
\usepackage[footnotesize,singlelinecheck=off,justification=raggedright]{caption}
\usepackage{multirow}
\def\bra#1{\mathinner{\langle{#1}|}}
\def\ket#1{\mathinner{|{#1}\rangle}}

\def\Bra#1{\left<1>}
\DeclareMathAlphabet{\mathbbmsl}{U}{bbm}{m}{sl}
{\catcode`\|=\active\gdef\Braket#1{\left<\mathcode`\|"8000\let|\bravert {#1}\right>}}
\def\bravert{\egroup\,\vrule\,\bgroup}
\def\Tr{\mathop{\mbox{\normalfont Tr}}\nolimits}

\def\ii{{\rm i}}
\def\ee{{\rm e}}
\begin{document}

\title{Entanglement entropy in the Long-Range Kitaev chain}

\author{Filiberto Ares\footnote{Corresponding author.}}
\email{ares@unizar.es}
\affiliation{Departamento de F\'{\i}sica Te\'orica, Universidad de Zaragoza,
50009 Zaragoza, Spain}
\author{Jos\'e G. Esteve}
 \email{esteve@unizar.es}
 \affiliation{Departamento de F\'{\i}sica Te\'orica, Universidad de Zaragoza,
50009 Zaragoza, Spain}
\affiliation{Instituto de Biocomputaci\'on y F\'{\i}sica de Sistemas
Complejos (BIFI), 50009 Zaragoza, Spain}
  \author{Fernando Falceto}
\email{falceto@unizar.es}
 \affiliation{Departamento de F\'{\i}sica Te\'orica, Universidad de Zaragoza,
50009 Zaragoza, Spain}
\affiliation{Instituto de Biocomputaci\'on y F\'{\i}sica de Sistemas
Complejos (BIFI), 50009 Zaragoza, Spain}

 \author{Amilcar R. de Queiroz}
 \email{amilcarq@gmail.com}
 
 \affiliation{Instituto de Fisica, Universidade de Brasilia, 
 Caixa Postal 04455, 70919-970, Bras\'{\i}lia, DF, Brazil}



\begin{abstract} 
  In this paper we complete the study on the asymptotic behaviour of
  the entanglement entropy for Kitaev chains with long
  range pairing. We discover that when the couplings decay
  with the distance with a critical exponent new properties
  for the asymptotic growth of the entropy appear.
  The coefficient of the leading
  term is not universal any more and the connection with conformal
  field theories is lost.  We perform a numerical and analytical
  approach to the problem showing a perfect agreement.
  In order to carry out the analytical study, a new technique
  for computing the asymptotic behaviour of block Toeplitz determinants
  with discontinuous symbols has been developed. 
\end{abstract}

\maketitle

\section{Introduction}

Entanglement plays a fundamental role in quantum phase transitions.
In one dimensional critical theories with local interactions, the
algebraic decay of the ground state correlation functions 
leads to the well-known logarithmic asymptotic growth of 
the entanglement entropy with a coefficient proportional 
to the central charge of the underlying conformal field theory 
\cite{Holzhey, Latorre, CarCal}. On the other hand, outside criticality, 
the correlations decay exponentially and the entanglement entropy satisfies 
an area law \cite{Hastings, Brandao}.  

The presence of long range interactions radically modifies the previous
picture \cite{Cirac, Koffel}. As it is discussed in
\cite{Vodola,Vodola2} correlations can display exponential and
algebraic decay even with non zero mass gap. This implies that the entanglement 
entropy may violate the area law while the system is non critical.
This happens in the so-called Long-Range Kitaev chain \cite{Vodola, Vodola2}, 
which has been the object of an intense study in the last years
\cite{Regemortel, Lepori, Patrick, Lepori2, Cirac2}.
Apart from the theoretical interest, as a model to analyse the role
of conformal symmetry in critical points, there has been recently
a spectacular development of experimental devices where these
long range interactions  can be simulated
\cite{Micheli, Islam, Britton, Richerme, Labuhn}. 

In \cite{Ares3} we studied the behaviour of the
entanglement entropy of the vacuum
with the size of the subsystem for a Long-Range
Kitaev chain. In order to compute the asymptotic behaviour of the entropy
we had to establish new results on block Toeplitz determinants
with discontinuous symbol. In that paper we obtained two different
regimes depending on the exponent $\zeta$ that governs the decay of the
couplings. The two regimes were already analysed numerically in
\cite{Vodola} and correspond respectively to $\zeta>1$ and $\zeta<1$.
In this paper we study the intermediate critical case $\zeta=1$ for
which new physical properties for the entropy show themselves. 
From the point of view of Toeplitz determinants, the critical exponent
posses new challenges that must be solved before being able
to determine the asymptotic behaviour of the entanglement entropy.
This task is carried out in the rest of the paper.

\section{The model}

The Long-Range Kitaev chain is a unidimensional homogeneous
fermionic chain with nearest-neighbour hopping and power-like 
decaying pairings, 
\begin{equation}\label{lrk}
H=\sum_{n=1}^N 
\Big( a_n^\dagger a_{n+1}+a_{n+1}^\dagger a_{n}+ h\,a_{n}^\dagger a_{n}
+\sum_{|l|<N/2} \frac{l}{|l|^{\zeta+1}} (a_{n}^\dagger a_{n+l}^\dagger - a_na_{n+l})
\Big)
-\frac{Nh}{2},
\end{equation}
where $a_n^\dagger$, $a_n$ are fermionic creation
and annihilation operators acting on the site $n$, 
satisfying the canonical anticommutation relations 
$\{a_n,a_m^\dagger\}=\delta_{nm}$, $\{a_n^\dagger,a_m^\dagger\}=
\{a_n,a_m\}=0$ and periodic boundary conditions, 
that is $a_{n+N}=a_n$.

The exponent $\zeta>0$ characterises the dumping of the coupling
with the distance. Its value plays a decisive role in
the growth rate of the R\'enyi entanglement entropy \cite{Vodola, Ares3, Vodola2}. 
In a previous work \cite{Ares3} we analytically studied the
leading asymptotic behaviour with the size of an interval
of the chain $|X|$ when $\zeta\neq 1$. There we found that 
the entanglement entropy $S_{\alpha,X}$ behaves as
\begin{equation}\label{scal_ee}
S_{\alpha,X}=\frac{\alpha+1}{6\alpha}c\log|X|+\dots
\end{equation}
up to finite contributions in the large $|X|$ limit, with
an effective central charge $c$ given by
\begin{equation}\label{eff_central_charge}
c=
\begin{cases}
  0&\zeta>1\ {\rm and}\ h\not=\pm2,\cr
  1/2&
  \zeta>1\ {\rm and}\ h=\pm2\ \ {\rm or}\ \
  \zeta<1\ {\rm and}\ h\not=2,\cr
  1&\zeta<1\ {\rm and}\ h=2.
\end{cases}
\end{equation}

This is in agreement with the numerical
studies previously performed in \cite{Vodola}.  

We see that the case $\zeta=1$ is the only one not
covered in the previous expression. To our knowledge
this case has not been considered in the literature
in spite of its physical interest as it may be experimentally
implemented with chains of magnetic impurities on an s-wave
superconductor \cite{Pientka}. Our reason to exclude this value
in \cite{Ares3} is that in order to address this case we required a
technical result that was not available at the moment
we wrote the previous paper. In this paper
we fill this gap and obtain the scaling behaviour
in the most general situation. We find that 
along the line $\zeta=1$ the entanglement entropy
grows logarithmically but with a proportionality
constant that can not be expressed in terms of an 
effective central charge. 
In the following we will develop the technical tools required
for the previous computation.

We must first solve the model in (\ref{lrk}).
Since it is a translational invariant and quadratic 
Hamiltonian, one can diagonalise it performing a Fourier
plus a Bogoliubov transformation (see e.g. Ref. \cite{Ares3}).
After the latter, the Hamiltonian reads
\begin{equation*}
  H=\sum_{k=0}^{N-1}\Lambda_k d_k^\dagger d_k,
\end{equation*}
where $d_k$ and $d_k^\dagger$ are also fermionic 
operators and $\Lambda_k$ is the dispersion relation
$$\Lambda_k=\sqrt{(h+2\cos\theta_k)^2+4\left(\sum_{l=1}^{N/2}
\sin(l\theta_k)l^{-\zeta}\right)^2},\quad 
\theta_k=\frac{2\pi k}{N}.$$
If we take now the thermodynamic limit $N\to \infty$,
$\theta_k$ is replaced by a continuous variable $\theta\in[-\pi, \pi)$ and
the dispersion relation can be expressed as
\begin{equation}\label{lrk_disp}
\Lambda(\theta)=\sqrt {(h+2\cos\theta)^2+|G_\zeta(\theta)|^2},
\end{equation}
where $G_\zeta(\theta)\equiv \Xi_\zeta({\rm e}^{{\ii}\theta})$ and
\begin{equation*}
  \Xi_\zeta(z)=\sum_{l=1}^\infty ({z^l-z^{-l}})~l^{-\zeta}=
  {\rm Li}_\zeta(z)-{\rm Li}_\zeta(z^{-1}).
\end{equation*}
The function ${\rm Li}_\zeta$ stands for the polylogarithm of order $\zeta$.
This is a multivalued function, analytic outside the real interval
$[1,\infty)$ and has a finite limit at $z=1$ for $\zeta>1$ while it 
diverges at that point for $\zeta<1$. For $\zeta=1$ the polylogarithm
function reduces to the logarithm ${\rm Li}_1(z)=-\log(1-z)$. 
We shall take as its branch cut the real interval $[1,\infty)$, 
thus $\Xi_1(z)$ has a branch cut along $[0,\infty)$. 
These properties will be crucial in the study of the 
entanglement entropy of the ground state of (\ref{lrk}). 

\begin{figure}[h]
  \centering
    \resizebox{12cm}{4cm}{\includegraphics{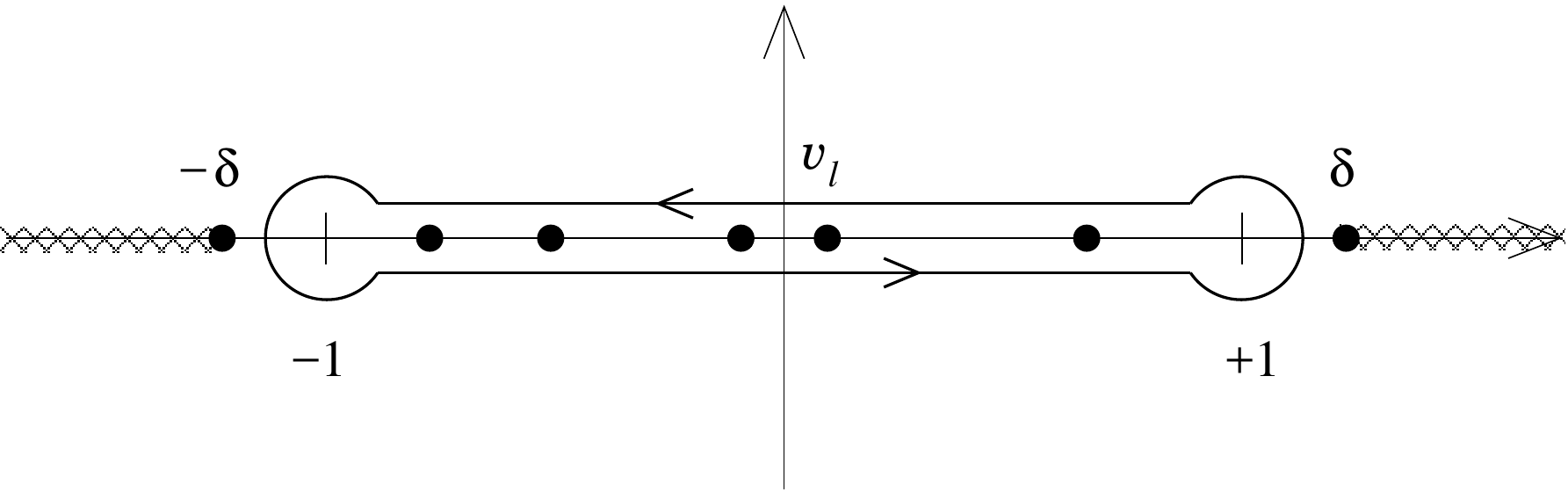}} 
    \caption{Contour of integration, cuts and poles for the computation of 
    $S_{\alpha, X}$. The contour surrounds the eigenvalues $v_l$ of $V_X$, all of
    them lying on the real interval $[-1,1]$. The cuts for the function $f_\alpha$ 
    extend to $\pm\infty$.}
  \label{contorno0}
   \end{figure}

Since $\Lambda_k\geq0$, the state of minimum energy $\ket{{\rm GS}}$
is the Fock space vacuum for the Bogoliubov modes, i. e.
$d_k\ket{\mathrm{GS}}=0$ for all $k$. 
If we split the system in two subchains $X$, $Y$ of contiguous
sites, the Hilbert space factorises
$\mathcal{H}=\mathcal{H}_X\otimes\mathcal{H}_Y$. Then the
R\'enyi entanglement entropy of $\ket{{\rm GS}}$ is defined as
\begin{equation*}
S_{\alpha,X}=\frac{1}{1-\alpha}\log\Tr(\rho_X^\alpha),
\end{equation*}
with $\rho_X=\Tr_{\mathcal{H}_Y}\ket{{\rm GS}}\bra{{\rm GS}}$
the reduced density matrix of $X$.
As it is well known, for Gaussian theories where the Wick theorem is 
satisfied, as it is our case, the entanglement entropy of this state
can be computed from the correlation matrix
\cite{Sorkin, Latorre, Peschel}. 
In fact \cite{Jin, Its, Its2}, 
\begin{equation}\label{entropy_contour}
S_{\alpha,X}=\lim_{\delta \to 1^+}\frac{1}{4\pi{\ii}}\oint_C f_\alpha(\lambda/\delta)
\frac{{\rm d}}{{\rm d}\lambda}\log D_X(\lambda){\rm d}\lambda,
\end{equation}
where 
\begin{equation}\label{falpha}
  f_\alpha(x)=\frac{1}{1-\alpha}\log\left[\left(\frac{1+x}{2}\right)^\alpha
    +\left(\frac{1-x}{2}\right)^\alpha\right],
\end{equation}
and $D_X(\lambda)=\det(\lambda I-V_X)$ with
$V_X$ the restriction of the ground state correlation matrix to the 
sites that belong to the subsystem $X$,
$$(V_X)_{nm}=\left\langle \left(\begin{array}{c} a_n \\ 
a_n^\dagger\end{array}\right)
\left(a_m, a_m^\dagger\right)\right\rangle-\delta_{nm}I, \quad n,m\in X.$$
The integration contour and the branch choice for $f_\alpha(\lambda/\delta)$
are represented in the Fig. \ref{contorno0}. 

Due to the translational invariance of the Hamiltonian
and given the choice of the subsystem $X$, which is composed of
contiguous sites, the
correlation matrix is a block Toeplitz matrix,
\begin{equation}
(V_X)_{nm}=\frac{1}{2\pi}\int_{-\pi}^{\pi}\mathcal{{G}}(\theta)
{\rm e}^{{\ii}\theta(n-m)}{\rm d}\theta,
\end{equation}
generated by a 2-dimensional symbol $\mathcal{{G}}(\theta)$ which,
for the ground state of the Hamiltonian (\ref{lrk}), has the form 
\begin{equation}\label{symbol}
\mathcal{{G}}(\theta)
=\frac1{\Lambda(\theta)}\left(\begin{array}{cc} h+2\cos\theta & 
G_\zeta(\theta) \\ 
-G_\zeta(\theta) & -h-2\cos\theta\end{array}\right).
\end{equation}

By virtue of (\ref{entropy_contour}), the asymptotic behaviour of $S_{\alpha, X}$ can 
be obtained from that of the determinant $D_X(\lambda)$. The discontinuities 
of the symbol (\ref{symbol}) will give the leading dominant term. 
In \cite{Ares3}, we already computed them for $\zeta\neq1$,
where the lateral limits of each discontinuity commute.
For $\zeta=1$ the symbol has a non-commuting discontinuity
and we lacked a way to determine its contribution. 
In the following we shall solve this problem.

\section{Determinant of block Toeplitz matrices with discontinuous symbol}

In this section we shall study the  scaling properties of the determinant
of a block Toeplitz matrix with piecewise continuous
matrix valued symbol.

Given a $d\times d$ dimensional matrix valued function
$\mathcal{M}(\theta)$ we shall denote by
$T_X[\mathcal{M}]$ its associated $|X|\!\cdot\! d$ dimensional
block Toeplitz matrix, i.e. for every $k,l=1,\dots,|X|$
there is a $d\times d$ block given by
$$\big(T_X[\mathcal{M}]\big)_{k,l}=
\frac1{2\pi}\int_{-\pi}^\pi\mathcal{M}(\theta){\rm e}^{\ii\theta(k-l)}{\rm d}\theta.$$
We shall denote by $D_X[\mathcal{M}]=\det T_X[\mathcal{M}]$.

From  Szego-Widom theorem \cite{Widom, Widom2, Widom3}
we know that the leading linear term of $\log D_X[\mathcal{M}]$
is given by
\begin{equation*}
\log D_X[\mathcal{M}]=\frac{|X|}{2\pi}\int_{-\pi}^\pi 
\log\det\mathcal{M}(\theta){\rm d}\theta+\cdots,
\end{equation*}
where we  assume $\det\mathcal{M}(\theta)\not=0$ and
the dots denote subdominant contributions in the limit $|X|\to\infty$.

As it is discussed in \cite{Ares3},
the next subdominant contribution 
for a piecewise continuous symbol with
jump discontinuities at $\theta_r$, for $r=1,\dots,R$, is logarithmic in $|X|$
and has the form
\begin{equation}\label{asymp_disc}
  \log  D_X[\mathcal{M}]=
  \frac{|X|}{2\pi}\int_{-\pi}^\pi 
\log\det \mathcal{M}(\theta){\rm d} \theta
+\log |X|\sum_{r=1}^R b_r+\cdots,
\end{equation}
where the coefficient $b_r$
depends uniquely on the lateral limits
at the discontinuity $\theta_r$.

In the particular case in which the two lateral limits commute
an explicit expression for $b_r$ is proposed
in \cite{Ares3}. We review here this result. 

If for a discontinuity $\theta_r$ the two lateral limits 
$$\mathcal{M}_r^-=\lim_{\theta\to\theta_r^-} \mathcal{M}(\theta)
\qquad{\rm and}\qquad 
\mathcal{M}_r^+=\lim_{\theta\to\theta_r^+}\mathcal{M}(\theta)
$$
commute and are diagonalisable, then we can diagonalise
both in the same basis.
Let us denote by $\mu_{r,j}^\pm,\ j=1,\dots,d$
the corresponding eigenvalues  for each lateral limit of the
discontinuity at $\theta_r$. Then
inspired by the similar case for a scalar symbol \cite{Fisher, BasorMorrison},
we conjectured in \cite{Ares3} that the logarithmic coefficients are given by
\begin{equation}\label{comm}
b_r=\frac{1}{4\pi^2} \sum_{j=1}^d\left(\log\frac{\mu_{r,j}^-}
{\mu_{r,j}^+}\right)^2.
\end{equation}
This conjecture was verified numerically to all attainable precision.
Note that  the previous expression can be written in the more compact
(and meaningful) form
\begin{equation}\label{commmatrix}
b_r=\frac{1}{4\pi^2}\Tr\left[\log\mathcal{M}_{r}^-
(\mathcal{M}_{r}^+)^{-1}\right]^2.
\end{equation}  
In the following we shall argue that if this result is true, it is also
valid when the two lateral limits do not commute.

In order to show it, we have to invoke an old result by Widom
\cite{Widom2} that, in the particular case that concerns our
derivation, can be stated in the very simple form:

{\it For any $d\times d$ constant matrix
$\mathcal{C}$ and any symbol $\mathcal{M}(\theta)$ as before we have
$$T_X[\mathcal{M}\mathcal{C}]=T_X[\mathcal{M}]T_X[\mathcal{C}].$$}
Actually $T_X[\mathcal{C}]=I_{|X|}\otimes\;\mathcal{C}$ and the relation above
immediately follows.
For the determinants we have
$$\log  D_X[\mathcal{M}\mathcal{C}]=\log D_X[\mathcal{M}]
+|X| \log\det \mathcal{C}.$$

Now, in the general case of a piecewise continuous symbol
with non commuting lateral limits, $\mathcal{M}_r^\pm$,
at the discontinuity point $\theta_r$, we can apply the
previous result choosing the constant symbol
$\mathcal{C}=(\mathcal{M}_r^+)^{-1}$. Hence we have
\begin{equation}\label{factor}
  \log  D_X[\mathcal{M}]=\log D_X[\mathcal{M}(\mathcal{M}_r^+)^{-1}]
  +|X| \log\det \mathcal{M}_r^+.
\end{equation}
Now, applying the expansion in (\ref{asymp_disc}) to the new symbol
$\mathcal{M}(\mathcal{M}_r^+)^{-1}$ we have
\begin{equation}\label{asymp_I}
\log D_X[\mathcal{M}(\mathcal{M}_r^+)^{-1}]=
\frac{|X|}{2\pi}\int_{-\pi}^\pi 
\log\det \mathcal{M}(\theta){\rm d} \theta
- |X|\log\det \mathcal{M}_r^+
+\log |X|\sum_{r=1}^R b'_r+\cdots.
\end{equation}
But the two lateral limits at $\theta_r$
are $\mathcal{M}_r^-(\mathcal{M}_r^+)^{-1}$ and $I$ which,
of course, commute. Therefore, we can apply the previous result
on commuting lateral symbols and we obtain
$$b'_r=\frac{1}{4\pi^2}\Tr\left[\log\mathcal{M}_{r}^-
  (\mathcal{M}_{r}^+)^{-1}\right]^2.$$
If we combine (\ref{asymp_disc}), (\ref{factor})
and (\ref{asymp_I}) we get $b_r=b'_r$ and therefore
(\ref{commmatrix}) is valid even if the lateral limits do not commute,
as stated.

\section{Entanglement entropy}

After the technical parenthesis of the previous section we
continue with the main goal of the paper.
Our task is to obtain the scaling behaviour of the entanglement entropy
$S_{\alpha,X}$ as derived from formula
(\ref{entropy_contour}).

For that purpose we must use the results of
the previous section to compute
the block Toeplitz determinant 
$D_X(\lambda)$
or, equivalently,
$D_X[\mathcal{G}_\lambda]$ 
in the notation of the last section.
Here we have introduced the 2-dimensional symbol
$\mathcal{G}_\lambda(\theta)=\lambda I-\mathcal{G}(\theta)$
with $\mathcal{G}(\theta)$ taken from (\ref{symbol}).

The linear term of $\log D_X(\lambda)$ is derived from 
the Szego estimate and, given that
$\det\mathcal{G}_\lambda(\theta)=\lambda^2-1$,
the asymptotic behaviour is
$$\log D_X(\lambda)=|X| \log(\lambda^2-1) + \log|X|\sum_r b_r(\lambda)+\dots,$$
where as before the dots represent finite terms in the large $|X|$ limit
and the coefficients of the logarithmic piece are associated
to the discontinuities 
of $\mathcal{G}_\lambda$
by the expression derived in the previous section.
Namely, the coefficient $b_r(\lambda)$ associated to the discontinuity at
$\theta_r$ is given by
$$b_r(\lambda)=\frac{1}{4\pi^2}\Tr\left[\log\mathcal{G}_{\lambda,r}^-
  (\mathcal{G}_{\lambda,r}^+)^{-1}\right]^2.
$$
Therefore, to proceed we must identify the location of the
jumps of $\mathcal{G}_\lambda$ and compute the lateral limits.

It is clear that the possible sources of discontinuities for
$\mathcal{G}_\lambda$ are the discontinuities or the zeros 
of $\Lambda(\theta)$. In our case of interest, $\zeta=1$,
these are $\theta=0$, for which $$G_{\zeta=1}(\theta)=\ii(\pi-\theta),
\ \ \theta\in[0,2\pi)$$ 
has a jump; and, for $h=2$, $\theta=\pi$ where $\Lambda(\theta)$ vanishes.
We will discuss separately both discontinuity points.

Let us call $b_0$ the coefficient in the expansion of
$\log D_X(\lambda)$ associated to $\theta=0$. The two lateral limits
are conveniently expressed in terms of the Pauli sigma matrices
$$\mathcal{G}_{\lambda,0}^\pm
=\lambda I-\cos\xi\ \sigma_z\pm\sin\xi\ \sigma_y,$$
where
$$\cos\xi=\frac{h+2}{\sqrt{(h+2)^2+\pi^2}},\quad
  \sin\xi=\frac{\pi}{\sqrt{(h+2)^2+\pi^2}}.
$$
Clearly for $\xi\not=\pi/2$ the two limits do not commute and we
are bound to use the results of the previous section.

With a little algebra we get
$$\mathcal{G}_{\lambda,0}^-(\mathcal{G}_{\lambda,0}^+)^{-1}=
\frac1{\lambda^2-1}\left(
(\lambda^2-\cos2\xi)I-2\lambda\sin\xi\ \sigma_y-\ii\sin\xi\ \sigma_x\right).
$$
In order to compute $b_0(\lambda)$ we need the eigenvalues
$\mu_\pm(\lambda)$ of the previous matrix.
They can be written
$$\mu_{\pm}(\lambda)=\left(\frac{\sqrt{\lambda^2-\cos^2\xi}\pm\sin\xi}
{\sqrt{\lambda^2-1}}\right)^2.$$
Notice also that we have
$$\mu_+(\lambda)=\mu_-(\lambda)^{-1}.$$
Putting everything together we finally obtain
\begin{eqnarray}
  b_0(\lambda)&=&\frac1{4\pi^2}\left[\left(
    \log\mu_+(\lambda)\,\right)^2+\left(\log\mu_-(\lambda)\,\right)^2\right]
    \cr
  &=&\frac2{\pi^2}\left(\log\frac{\sqrt{\lambda^2-\cos^2\xi}+\sin\xi}
  {\sqrt{\lambda^2-1}}\right)^2.
  \end{eqnarray}

As we discussed before, for $h\not=2$ this is the only discontinuity
of the symbol and therefore the only contribution to the logarithmic
term for  $\log D_X(\lambda)$. Namely
$$\log D_X(\lambda)=|X| \log(\lambda^2-1) + \log|X| b_0(\lambda)+\dots.$$
This must be inserted into (\ref{entropy_contour}) to derive the
scaling behaviour of the entanglement entropy.

The linear term is obtained from the integral
$$\lim_{\delta \to 1^+}\frac{1}{2\pi{\ii}}\oint_C f_\alpha(\lambda/\delta)
\frac{{\lambda}}{\lambda^2-1}{\rm d}\lambda=\frac12
(f_\alpha(1)+f_\alpha(-1))=0,$$
where the Cauchy's residue theorem has been used and from the expression
in (\ref{falpha}) we immediately see that $f_\alpha(\pm1)=0$.

The previous result: vanishing of the term that scales linearly
with the size of the subsystem,
is in agreement with the {\it area law}, that states that
in the ground state of a local theory the entanglement entropy is not
an extensive quantity. As we will see now the area law has corrections
and the entanglement entropy scales with the logarithm
of the size of the subsystem rather than going to a constant.

\begin{figure}[h]
  \centering
    \resizebox{12cm}{4cm}{\includegraphics{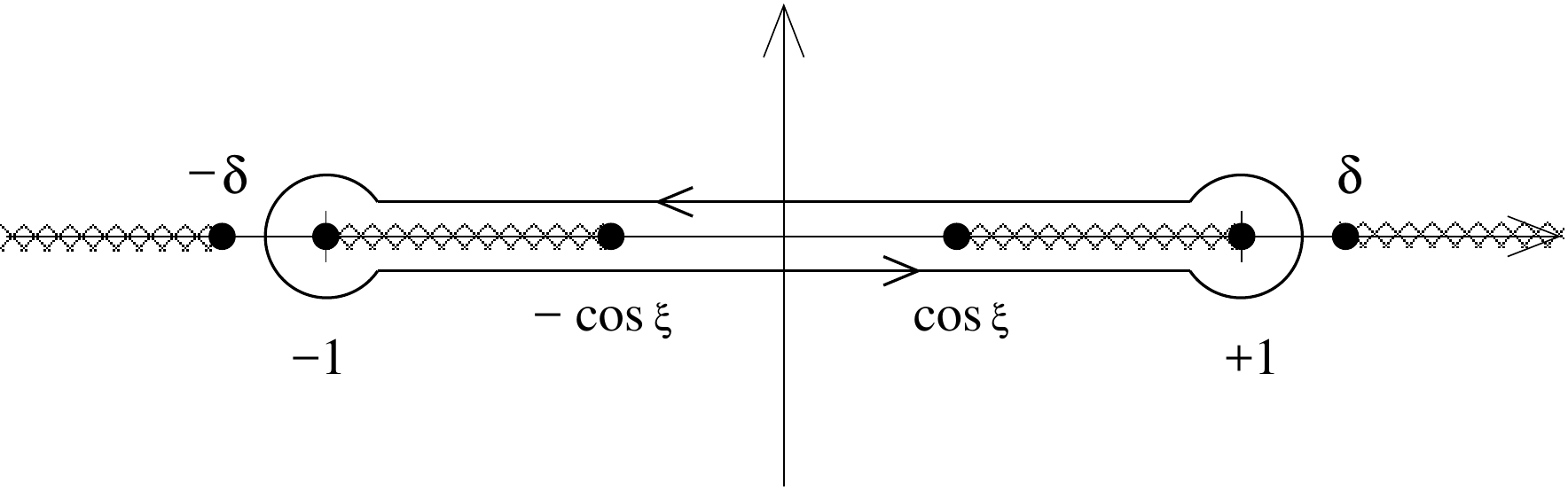}} 
    \caption{Contour of integration and cuts of the integrand in (\ref{intB_alpha})
    for the computation of $B_{\alpha,0}$. The cuts from $\pm \delta$ to the infinity
    correspond to ${\rm d}f_\alpha(\lambda/\delta)/{\rm d}\lambda$ while the cuts inside
    the contour, $[-1,-\cos\xi]$ and $[\cos\xi,1]$, are due to the other factor of the 
    integrand.} 
  \label{contorno1}
   \end{figure}

The coefficient of the logarithmic term for the entropy,
that we denote by $B_\alpha$, gets just one contribution,
named $B_{\alpha,0}$,  from the discontinuity at $\theta=0$;
we are considering $h\not=2$.
It can be computed from (\ref{entropy_contour})
\begin{eqnarray}\label{intB_alpha}
B_{\alpha,0}&=&\lim_{\delta \to 1^+}\frac{1}{4\pi{\ii}}\oint_C f_\alpha(\lambda/\delta)
\frac{{\rm d}b_0(\lambda)}{{\rm d}\lambda} {\rm d}\lambda\cr
&=&
-\lim_{\delta \to 1^+}\frac{1}{2\pi^3{\ii}}\oint_C
\frac{{\rm d} f_\alpha(\lambda/\delta)}{{\rm d}\lambda}
\left(\log\frac{\sqrt{\lambda^2-\cos^2\xi}+\sin\xi}
  {\sqrt{\lambda^2-1}}\right)^2{\rm d}\lambda,
\end{eqnarray}
where, for convenience, we have performed an integration by parts.
The branch cuts of the different multivalued functions
involved in the integral are depicted in the Fig. \ref{contorno1}.

Now it is possible to take the limit in (\ref{intB_alpha})
and, given that $f_\alpha$ is an even function, the complex integral can
be reduced to the following real one
\begin{equation}\label{b0}
B_{\alpha,0}=
\frac{2}{\pi^2}\int_{\cos \xi}^1
\frac{{\rm d} f_\alpha(\lambda)}{{\rm d}\lambda}
\log\frac{\sqrt{1-\lambda^2}}{\sqrt{\lambda^2-\cos^2\xi}+\sin\xi}
         {\rm d}\lambda,
\end{equation}
where we take positive square roots.

Note that for integer $\alpha>1$
\begin{equation}
\frac{{\rm d}f_\alpha(\lambda)}{{\rm d}\lambda}=\frac{\alpha}{1-\alpha}
\frac{(1+\lambda)^{\alpha-1}-(1-\lambda)^{\alpha-1}}{(1+\lambda)^\alpha+
(1-\lambda)^\alpha},
\end{equation}
is a meromorphic function
with poles along the imaginary axis
located at
$$\lambda_k= {\ii}\tan\frac{\pi(2k-1)}{2\alpha},
\quad k=1,2,\dots,\alpha,\ {\rm with}\ k\not=\frac{\alpha+1}2.$$
Hence, by sending the integration contour in (\ref{intB_alpha})
to infinity we can reduce the integral
to the computation of the corresponding residues.
In this way we get a completely explicit expression
for $B_{\alpha,0}$ (valid only for integer $\alpha>1$)
$$B_{\alpha,0}=\frac1{\pi^2(\alpha-1)}\sum_{k=1}^{\alpha}
\arctan^2\frac{\sin\xi}{\sqrt{\cos^2\xi+|\lambda_k|^2}}.
$$
In particular, for $\alpha=2, 3$ the sum in the previous expression
reduces to
\begin{eqnarray*}
  B_{2,0}&=&\frac2{\pi^2}
  \arctan^2\frac{\sin\xi}{\sqrt{\cos^2\xi+1}},
  \\
 B_{3,0}&=&\frac1{\pi^2}
  \arctan^2\frac{\sin\xi}{\sqrt{\cos^2\xi+1/3}}.
\end{eqnarray*}

\begin{figure}[h]
  \centering
    \resizebox{15.75cm}{10cm}{\includegraphics{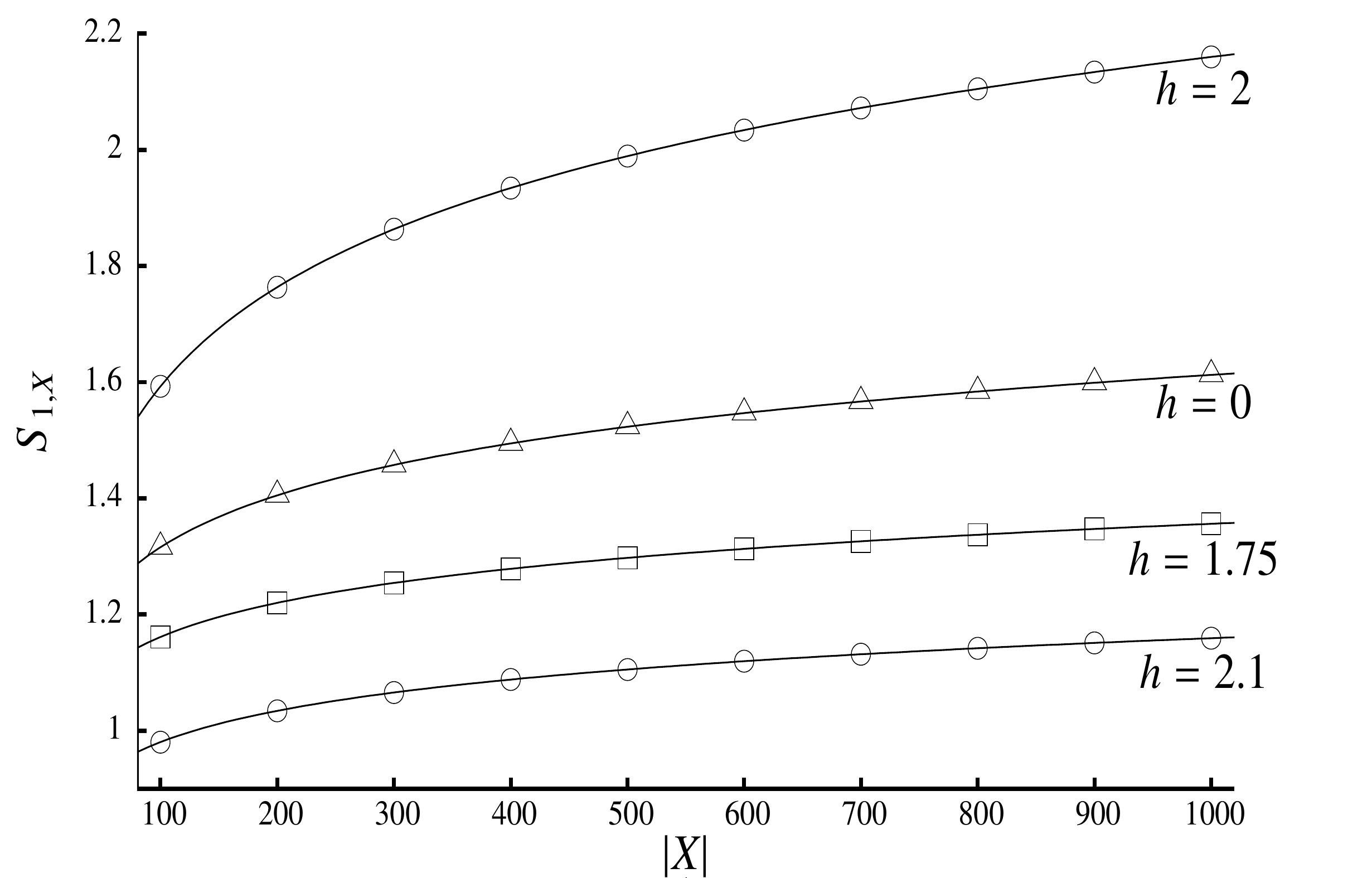}} 
    \caption{Numerical check of the asymptotic scaling of the von Neumann entanglement
    entropy ($\alpha=1$) with the length of the interval $|X|$ for $\zeta=1$ and
    different values of $h$. The dots represent the numerical computation
    while the solid lines correspond to the curve $B_1\log|X|+{\rm constant}$ where
    $B_{1}=B_{1,0}$ for $h\neq 2$ and $B_{1}=B_{1,0}+1/6$
    if $h=2$. $B_{1,0}$ is given by (\ref{b0}).
    The constant is obtained by fitting the numerical and the analytical
    results at $|X|=1000$.}
  \label{numerics_lrk}
   \end{figure}

In the Fig. \ref{numerics_lrk} we numerically check the scaling
of the entanglement entropy given by the expression (\ref{b0})
that we have just obtained for the logarithmic coefficient $B_\alpha$. 
Although we have only plotted the von Neumann entropy, we have seen 
that the agreement with the numerical computations is just as good 
for other values of $\alpha$. 

For local theories it is possible to relate the entanglement entropy
to universal properties of conformal field theories. In this case
the coefficient of the logarithmic term in the expansion
should read
\begin{equation}\label{central_charge}
B_\alpha=
\frac{\alpha+1}{6\alpha}c,
\end{equation}
where $c$, the central charge, is a constant that depends
on the universality class of the underlying conformal field theory.
In particular this relation implies
$$B_2=\frac98 B_3,$$
something that, for general $h$, does not hold in our case.
The reason for that could be that couplings in our theory have
infinite range and therefore it cannot be related to any local field theory. 
However, as it was shown in \cite{Ares3}, when the coupling
constants decay faster ($\zeta>1$) or
slower ($\zeta<1$) than our critical case, relation
(\ref{central_charge}) also holds, even if the theory
is not local. Another instance in which $(\ref{central_charge})$
holds is when $h=-2$, then $\xi=\pi/2$ and we have
$$B_{2,0}=\frac18,\qquad
B_{3,0}=\frac1{9}, 
$$
which fulfil (\ref{central_charge}) with $c=1/2$.

This is the whole story about the logarithmic coefficient, and
$B_\alpha=B_{\alpha,0}$ for $h\not=2$, as we discussed before.
However for $h=2$ the dispersion relation $\Lambda(\theta)$ vanishes at
$\theta=\pi$ and it induces a discontinuity in the symbol.
The two lateral limits are
$$\mathcal{G}_{\lambda,\pi}^\pm=\lambda I\pm\sigma_y$$
and its contribution to the logarithmic
term in the expansion of $\log D_X(\lambda)$
is given by
\begin{eqnarray}
b_\pi(\lambda)&=&
\frac{1}{4\pi^2}\Tr\left[\log\mathcal{G}_{\lambda,\pi}^-
  (\mathcal{G}_{\lambda,\pi}^+)^{-1}\right]^2\cr
&=&\frac{1}{2\pi^2}\left(\log\frac{\lambda+1}{\lambda-1}
\right)^2.
\end{eqnarray}
Associated to that there is a new logarithmic term for the entropy
whose coefficient is given by
\begin{equation}\label{bpi}
B_{\alpha,\pi}=
\lim_{\delta\to1^+}\frac{1}{4\pi{\ii}} \oint_C f_\alpha(\lambda/\delta) 
\frac{{\rm d}}{{\rm d}\lambda}b_{{\pi}}(\lambda){\rm d}\lambda
=\frac{\alpha+1}{12\alpha}.
\end{equation}
In the light of the previous discussion this contribution
to the logarithmic term can be interpreted as coming from a
conformal field theory with central charge $c=1/2$.

We would like to stress again that the last interpretation is valid only for
part of the contribution to the logarithmic term of the entropy.
One should consider the whole coefficient $B_\alpha=B_{\alpha,0}+
B_{\alpha,\pi}$, that includes the discontinuity at
$\theta=0$, and this, for $h=2$, does not behave
as predicted by the conformal field theory under variations of $\alpha$.
Observe that in the Fig. \ref{numerics_lrk} we have considered this particular
point. Also in this case the analytical formulae we have obtained are in agreement
with the numerical results. 

\begin{figure}[H]
  \centering
{\includegraphics{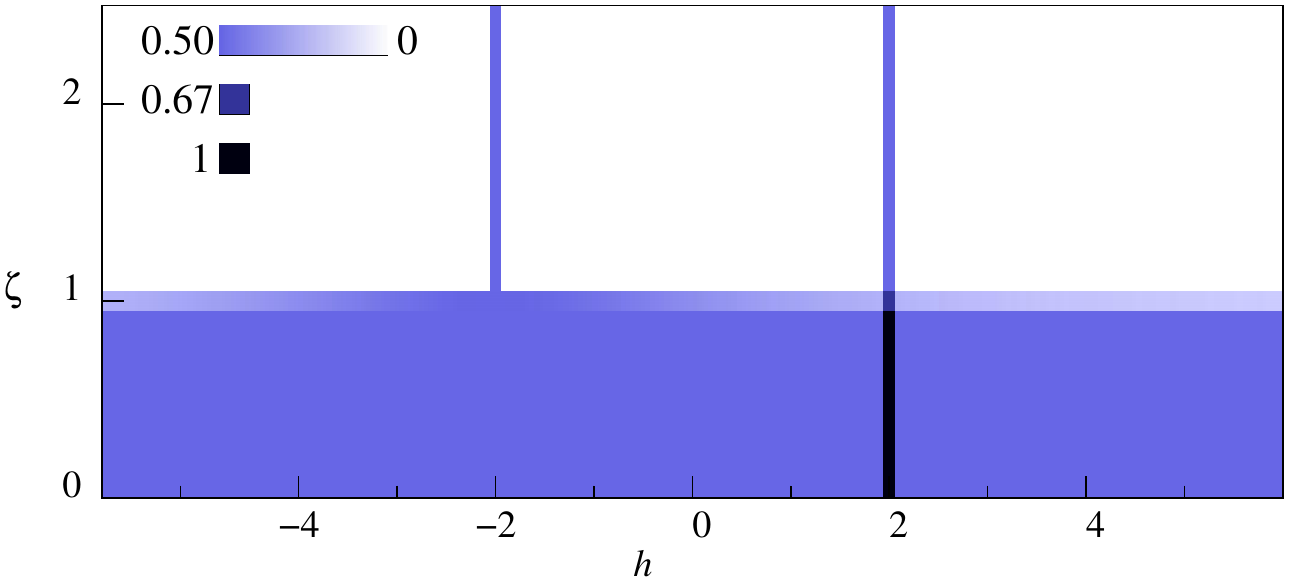}} 
    \caption{
Plot of the different regions in the $(h,\zeta)$ plane according to the 
coefficient $B_2$ of the logarithmic scaling for the entanglement entropy. The 
colouring stands for the value of $2B_2$: from white for $B_2=0$ to black when $B_2=1/2$. 
}
  \label{kitaev2}
   \end{figure}

As a summary, in the Fig. \ref{kitaev2} we have recollected the previous known
results (\ref{scal_ee}), (\ref{eff_central_charge}) outside the line $\zeta=1$ 
together with those obtained in this section for $\zeta=1$. We have coloured the 
parameter space $(h,\zeta)$ according to the value of the logarithmic term $B_\alpha$.
The model is critical, the mass gap is zero, 
only in the lines $h=\pm 2$. Note, however, that the entanglement entropy scales logarithmically 
for $\zeta\leq 1$ even outside criticality.

As a further check of the results obtained in the previous
section we shall consider a variant of the Long-Range Kitaev
Chain. Namely we shall discuss the case in which we have
two singularities in the symbol (instead of one) located at $\theta=\pm\phi$.
One reason for this is connected to the study of the
M\"obius symmetry introduced in \cite{Ares2,Ares4}.
While $\theta=0$ is a fixed point of the above mentioned transformations
$\theta=\pm\phi$ is not, which enriches the symmetry.
We do not pursue the analysis of M\"obius transformations
in this paper.

The new singularities in the symbol can be obtained by adding 
to (\ref{lrk}) an oscillatory factor in the pairing,
\begin{equation}\label{lrk2}
H'=\sum_{n=1}^N \Big( a_n^\dagger a_{n+1}+a_{n+1}^\dagger a_{n}+ 
h\,a_{n}^\dagger a_{n}
+\sum_{|l|<N/2} \frac{l\cos(l\phi)}{|l|^{\zeta+1}} 
(a_{n}^\dagger a_{n+l}^\dagger - a_na_{n+l})\Big) 
-\frac{Nh}{2},
\end{equation}
with $\phi\in[0,\pi)$.  

In the thermodynamic limit $N\to\infty$, the dispersion relation
is
\begin{equation*}
\Lambda'(\theta)=\sqrt{(h+2\cos\theta)^2+|G_{\zeta,\phi}(\theta)|^2},
\end{equation*}
where $G_{\zeta,\phi}(\theta)=\Xi_{\zeta,\phi}({\rm e}^{{\ii}\theta})$, and
\begin{eqnarray*}
G_{\zeta,\phi}(\theta)&=&
\sum_{l=1}^\infty\frac{\cos(l\phi)}{l^\zeta}
(\ee^{{\ii}\,\theta l}-\ee^{-{\ii}\,\theta l})\\
&=&\frac{1}{2}\left[{\rm Li}_\zeta({\rm e}^{{\ii}(\phi+\theta)})-
{\rm Li}_\zeta({\rm e}^{{\ii}(\phi-\theta)})+{\rm Li}_\zeta({\rm e}^{-{\ii}(\phi-\theta)})-
{\rm Li}_\zeta({\rm e}^{-{\ii}(\phi+\theta)})\right].
\end{eqnarray*}

This function  vanishes at $\theta=0$ and $\pi$,
it is smooth for $\zeta>1$, diverges at $\theta=\pm\phi$ for $\zeta<1$
and for our particular case of interest in this paper, $\zeta=1$, it reads
$$
G_{1,\phi}(\theta)=
  \left\{\begin{array}{cc}-{\ii}(\pi+\theta),\ & -\pi\leq \theta \leq -\phi,\\
  -{\ii}\theta, & -\phi<\theta<\phi, \\
  {\ii}(\pi-\theta), & \phi\leq \theta <\pi.
  \end{array}\right.
$$
It has a jump at $\theta=\phi$ with lateral limits 
$\ii\phi$ and $\ii(\pi-\phi)$ and another one at
$\theta=-\phi$ with limits $\ii(\phi-\pi)$
and $-\ii\phi$.

The symbol of the correlation matrix can be written
$$\mathcal{{G}}'(\theta)
=\frac1{\Lambda'(\theta)}\left(\begin{array}{cc} 
h+2\cos\theta & G_{\zeta,\phi}(\theta)
\\ -G_{\zeta,\phi}(\theta) & -h-2\cos\theta\end{array}\right).$$

For $h\not=\pm2$ and $\zeta=1$ the only jumps
of $\mathcal{G}_\lambda'(\theta)=\lambda I-\mathcal{{G}}'(\theta)$
take place at $\theta=\pm\phi$. We shall compute separately
the contribution of each discontinuity to the coefficient  
of the logarithmic term in the expansion of the entanglement entropy.

In first place we consider the point $\theta=\phi$.
The lateral limits of  $\mathcal{G}_\lambda'(\theta)$ are
$$\mathcal{G}_{\lambda,\phi}^{\prime\pm}=\lambda I-\cos\xi^\pm\;\sigma_z-\sin\xi^\pm\,\sigma_y,$$
where
$$
\cos\xi^+=\frac{h+2\cos\phi}{\sqrt{(h+2\cos\phi)^2+(\phi-\pi)^2}},\quad
\sin\xi^+=\frac{\phi-\pi}{\sqrt{(h+2\cos\phi)^2+(\phi-\pi)^2}}
$$
and
$$
\cos\xi^-=\frac{h+2\cos\phi}{\sqrt{(h+2\cos\phi)^2+\phi^2}},
\quad\quad\quad\ 
\sin\xi^-=\frac{\phi}{\sqrt{(h+2\cos\phi)^2+\phi^2}}.\quad\quad\ $$
The eigenvalues of
$\mathcal{G}_{\lambda,\phi}^{\prime-}(\mathcal{G}_{\lambda,\phi}^{\prime+})^{-1}$
are:
$$\mu'_{\pm}(\lambda)=\left(\frac{\sqrt{\lambda^2-\cos^2(\Delta\xi/2)}\pm\sin(\Delta\xi/2)}
{\sqrt{\lambda^2-1}}\right)^2,$$
where $\Delta\xi=\xi^+-\xi^-$.

Observe that the expression for the eigenvalues is exactly that of
the model in the previous section, with the only change of $\xi$ by
$\Delta\xi/2$.

\begin{figure}[H]
  \centering
    \resizebox{15.75cm}{10cm}{\includegraphics{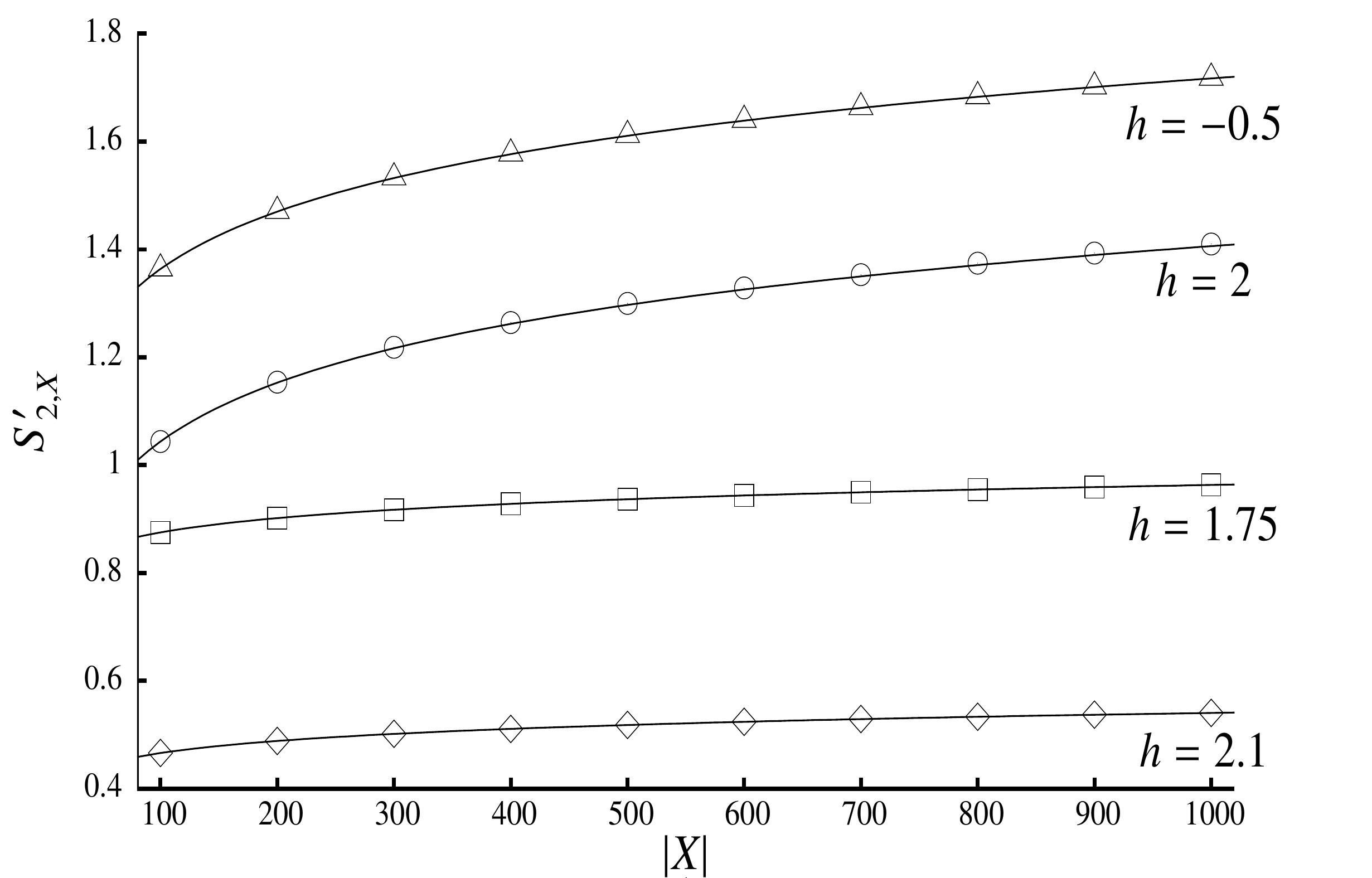}} 
    \caption{R\'enyi entanglement entropy with $\alpha=2$ as a function
    of the length $|X|$ of the interval for $\phi=\pi/4$, $\zeta=1$ and 
    different values of $h$. The dots correspond to the numerical entropy
    while the solid lines are our analytical prediction $B'_2\log|X|+{\rm constant}$ 
    with $B'_2$ that obtained in the text: $B'_2=2B'_{2,\pi/4}+1/4$ for $h=\pm2$ and
    $B'_2=2B'_{2,\pi/4}$ with $B_{2,\pi/4}$ given by (\ref{bphiint}).
    The constant is obtained by fitting the numerical and the analytical
    results at $|X|=1000$.}
  \label{numerics_lrk_mod}
   \end{figure}

Therefore the contribution of this discontinuity
to the logarithmic coefficient of the entanglement entropy,
$B'_{\alpha,\phi}$, can be written
\begin{equation}\label{bphi}
B'_{\alpha,\phi}=
\frac{2}{\pi^2}\int_{\cos\frac{\Delta\xi}2}^1
\frac{{\rm d} f_\alpha(\lambda)}{{\rm d}\lambda}
\log\frac{\sqrt{1-\lambda^2}}{\sqrt{\lambda^2-\cos^2(\Delta\xi/2)}+
  \sin(\Delta\xi/2)}
         {\rm d}\lambda,
\end{equation}
and the corresponding integrated expression for integer $\alpha>1$
\begin{equation}\label{bphiint}
B'_{\alpha,\phi}=\frac{1}{\pi^2(\alpha-1)}\sum_{k=1}^\alpha 
\arctan^2\frac{\sin(\Delta\xi/2)}{\sqrt{\cos^2(\Delta\xi/2)+|\lambda_k|^2}}.
\end{equation}
The other discontinuity point, $\theta=-\phi$, can be computed along
the same lines.
Actually the only difference with respect to the previous case is that we have to
replace $\xi^+$ and $\xi^-$ with $-\xi^-$ and $-\xi^+$ respectively.
Hence, $\Delta\xi$ is unchanged, $B'_{\alpha,-\phi}=B'_{\alpha,\phi}$
and for $h\not=\pm2$
the asymptotic behaviour of the entanglement entropy
for this modified Long-Range Kitaev chain is
$$S'_{\alpha,X}=2B'_{\alpha,\phi}\log|X|+\dots,$$
with 
$B'_{\alpha,\phi}$
given above. In the Fig. \ref{numerics_lrk_mod} we compare
the numerical value of the entanglement entropy for different
values of $h$ along the line $\zeta=1$ when $\alpha=2$ and 
$\phi=\pi/4$. The agreement between them makes us conclude
that the expression that we have obtained for the leading
contribution of the discontinuities of a block Toeplitz matrix
is correct.

For completeness we also consider the case $h=\pm2$ that was excluded
in the previous considerations. The Hamiltonian
in this case is gap-less
and the coefficient in the expansion of the entanglement entropy
gets an extra coefficient due to this fact.
In both cases $h=2$ or $-2$ the extra term is the same and
coincides with the one we obtained in the previous section.
Namely, for $h=\pm2$ we have
$$B_\alpha'=2B'_{\alpha,\phi}+ \frac{\alpha+1}{12\alpha}.$$
In the Fig. \ref{numerics_lrk_mod} we also checked the case 
$h=2$ when $\alpha=2$. Observe that the numerical values agree 
with the form of $B'_\alpha$ deduced for this case.

\section{Conclusions}

 In this work, we have completed the study of the scaling 
 behaviour of the ground state entanglement entropy 
 in the Long-Range Kitaev chain that we started in 
 \cite{Ares3}. Our analysis is based on the relation 
 between the entanglement entropy and the determinant 
 of the correlation matrix. Since the chain is translational 
 invariant, the correlation matrix for a single interval 
 is block Toeplitz. 

 If the symbol of the block Toeplitz matrix is continuous, the entropy
 scales with the area of the subsystem and hence has a finite asymptotic
 limit. On the contrary, the presence of discontinuities
 gives rise to corrections to the area law,
 leading to a logarithmic growth of the entropy with the size 
 of the interval. In \cite{Ares3} we considered the case $\zeta\neq 1$. 
 There, we  were able to compute the asymptotic behaviour
 of the entanglement entropy thanks to the fact that the lateral limits
 of the jump discontinuities commute in that case.
 However, for the intermediate value $\zeta=1$ we have jump discontinuities whose
 lateral limits do not commute and the results of \cite{Ares3} cannot be
 applied. Motivated by this physical problem, we have investigated more deeply
 into the theory of block Toeplitz determinants with discontinuous symbols
 and finally we have produced a general expression for the 
 leading contribution to the determinant of the discontinuities
 (both commutative and non-commutative) in the matrix symbol.
 To our knowledge, this is a new result. We have checked it numerically
 and we find a complete agreement with the theoretical predictions.

 A striking feature of the entanglement entropy
 in the intermediate case is that it cannot be derived from a conformal field
 theory. Actually in the non critical case, $\zeta\not=1$,
 the coefficient of the leading logarithmic term of the R\'enyi
 entanglement entropy can be derived from a conformal field theory thanks
 to the replica trick. This fixes the dependence of the coefficient
 on the R\'enyi exponent and the only free parameter corresponds to
 the effective central charge of the underlying conformal theory.
 On the contrary, in the intermediate case, $\zeta=1$, the logarithmic terms
 originated in  non-commutative discontinuities show a different
 dependence on the R\'enyi exponent and therefore can not be related
 to a conformal field theory.
 
 It would be nice to understand better this property. A starting point
 could be the analysis performed in \cite{Vodola, Vodola2} on the asymptotic
 form of the correlations for fermionic chains with long range pairing.
  \newline

\noindent{\bf Acknowledgements:} Research partially supported by grants 
2016-E24/2, DGIID-DGA and FPA2015-65745-P, MINECO (Spain). 
FA is supported by FPI Fellowship No. C070/2014, DGIID-DGA/European Social Fund.

 \end{document}